\documentclass[aps,graphicx,preprint]{revtex4}
\usepackage{amsmath}
\usepackage{amssymb}
\usepackage{graphicx}
\usepackage{amssymb}
\usepackage{subfigure}
\usepackage{pdfpages}
\usepackage{CJK}

\begin{document}

\title{Effective mass sensing using optomechanically induced transparency in microresonator system}

\author{Yong-Pan Gao$^{1,2}$, Tie-Jun Wang$^{1,2,5}$, Cong Cao$^{1,3}$, Si-Chen Mi$^{1,2}$, Daquan Yang$^{1,4}$, Yong Zhang$^{1,2}$, and Chuan Wang$^{1,2,*}$}

\address{$^1$ State Key Laboratory of Information photonics and Optical Communications, Beijing University of Posts and Telecommunications, Beijing, 100876,China\\
$^2$  School of Science, Beijing University of Posts and Telecommunications, Beijing, 100876,China\\
$^3$  School of Ethnic Minority Education, Beijing University of Posts and Telecommunications, Beijing, 100876, China\\
$^4$  School of Information and Communication Engineering, Beijing University of Posts and Telecommunications, Beijing 100876, China\\
$^5$ wangtiejun@bupt.edu.cn\\
$^*$ wangchuan@bupt.edu.cn} 

\begin{abstract}
Detecting and weighing the individual nanoparticles is an important approach to study the behavior and properties of single particles. Here we illustrate an effective mass sensing scheme using optomechanical resonator system. Based on the optomechanically induced transparency phenomenon, a Stokes field reference approach is used to sense the mass of the particle on the microresonator. The field intensity of the transmission field will be changed by the effect of the particle which avoids the limits of decay induced spectral width in the resonance shift detection. Exploiting the perturbation method, we theoretically evaluated the dynamical behavior of the system, and achieved the femtogram level mass sensing without the need for high cavity $Q$-value, and strong coupling strength in the optomechanically system.
\end{abstract}

\maketitle



\section{Introduction}

Mass sensing, which is used to weigh the external nanoparticle and molecules with tiny mass, has attracted much research interest. To date, most of the mass sensing schemes are based on electrical or optical techniques. For example, the electical schemes are usually based on the nano-electromechanical system(NEMS) [1-4]. Especially, the NEMS is generally constructed by a sorbent coated cantilever or bridge, when a single particle is deposited on the system, it would induce an intrinsic frequency shift result from the change of the effective mass. On the basis of this frequency shift, the mass change induced by the nanoparticle can be read out. To our knowledge, the precision of the electrical sensing scheme is approaching the attogram level in vacuum [5]. However, the ohmic effect and other intrinsic energy loss in the electric circuitry will broaden the electrical response spectrum and finally affect the sensitivity of the mass detection during the frequency measurement. Compared with the traditional electrical methods, optical sensors do not require on-chip power and allows for remote sensing via free-space or optical fibers [6]. Especially by exploiting the whispering-gallery-mode (WGM) resonators as the platform, various schemes of particle detection have realized. For example, Zhu \emph{et al.} realized the single nanoparticle sensing using silica optical microtoroids in 2010 [7]. Later exploiting the Erbium doped microcavities, He \emph{et al.} investigated the single viruses and nanoparticles sensing using whispering gallery mode microlaser [8]. Further by using Raman effect, the sensitivity and precision are improved [9,10]. During the past decades, various important progress on this field has been made [11-14]. Also, with the development of cavity optomechanics, the optical pump-probe response becomes a popular issue in the study of optomechanical system [15-17]. Exploiting the mechanical system, various studies and applications in optical and electrical systems have been presented, such as the optomechanics [18-20], nanomechanics [21-23], and so on.

The mechanical resonator based mass sensing relies on the resonant frequency-shift due to the increment of the mass value. In the past decades, there are various experiment progress on the branches of particle sensing and mass sensing using optical and mechanical systems. In 2013, Chaste \emph{et al.} experimentally investigate the mass sensing using carbon nanotubes with a resolution of $10^{-24} g$ [24]. In nanomechanical systems, Jensen \emph{et al.} [25] demonstrated a nanoscale mechanical resonator for atomic mass sensing with sensitivity of $1.3\times 10^{-25} kg/ \sqrt{Hz}$. The mass sensor operated in the nonlinear optical domain was demonstrated in the toroidal nanocavity optomechanically system [26]. Later, the sub-picogram  mass sensing and measurement with an optomechanical oscillator was experimentally presented [27]. Then, sensors using micro fluidic optomechanical resonators at the picogram level in liquid environment was experiment achieved [28]. Recently, a coupled double-cavity optomechanical system was also theoretical exploited to sense the mass at picogram level [29]. It is worthwhile to note that these schemes have achieved mass sensing based on the direct frequency shift detection, which means that the accuracy of the system will be limited by the spectra linewidth which could be called as inter-band sensing.

In this study, we propose an efficient mass sensing scheme to weigh the external nanoparticles deposited onto the surface of a mechanical resonator, by using pump-probe response of the optomechanical system under the condition of optomechanically induced transparency (OMIT) [30-34]. OMIT is the analogue of electromagnetically induced transparency (EIT) [35,36] which is induced by optomechanics. In the pump-probe driving progress of cavity optomechanical system under the OMIT condition, the four-wave mixing (FWM) [37-39] would produce the stokes field, which shows different transparency characteristics compared with the OMIT field when the mechanical resonator exhibits a slight frequency change.

We investigate this effect theoretically and conclude that the variance of the output fields intensity rather than the frequencies, also we obtain the change of the intensity with the increment of mass on the mechanical resonator. Compared with the previous schemes based on the detection of frequency shift, the present scheme avoids the effect of the cross-talk in the spectrum. We can therefore read out the frequency even if the full width half maximum (FWHM) far overweight than the frequency shift. Moreover, the most important issue in the destiny detection is the unavoidable decay in the propagation of the output signal.



\section{Model and Hamiltonian} \label{model}

\begin{figure}
\centering
\includegraphics[width=0.76\textwidth]{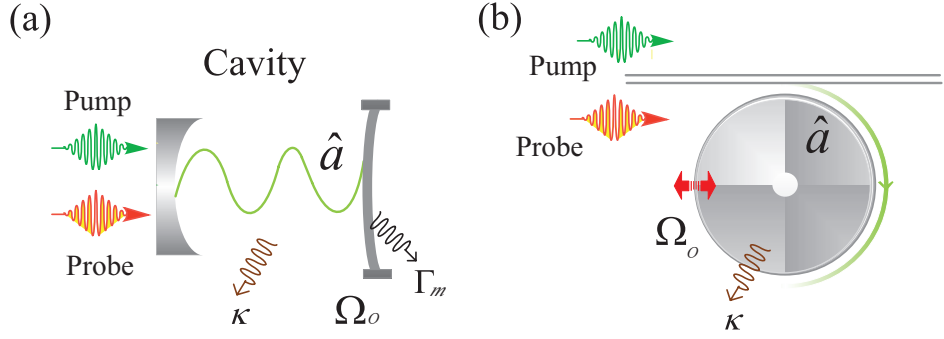}
\caption{(color online) Schematic diagram of a generic Fabry-Perot and whispering-gallery-mode optomechanical system. The optomechanical system is driven by a strong pump field with frequency $\omega_{1}$. with a weak probe field $\omega_{p}$,  many strange optical effect occur, optomechanically induced transparency. The intrinsic loss of the cavity is denoted as $\kappa$. The coupling strength between the waveguide and cavity is $\kappa_{ex}$. }
\label{picture}
\end{figure}

Traditionally, the optomechanical system consists of a Fabry-Perot cavity with one side is movable with the mechanical oscillator as shown in Fig.\ref{picture}(a). Similarly, as shown in Fig.\ref{picture}(b), the system could also be realized by WGM microresonator which supports both an optical mode and an optomechanical mode. When the system is coherently driven by both the pump and probe field, the Hamiltonian can be written as $\hat{H} =\hat{H}_{mech}+\hat{H}_{opt}+\hat{H}_{int}+\hat{H}_{dirve}$. Here $\hat{H}_{mech}$ and $\hat{H}_{opt}$ represent the Hamiltonian of the optomechanical field and the cavity field, respectively, which could be described as
\begin{eqnarray}
& \hat{H}_{mech} =\frac{\hat{p}^{2}}{2m_{eff}}+\frac{1}{2}m_{eff}\Omega_{m}^{2}\hat{x}^{2},\\
& \hat{H}_{opt} =\hbar \omega_{c}(\hat{a}^{\dag}\hat{a}+\frac{1}{2}),
\end{eqnarray}
where $\hat{p}$ and $\hat{x}$ are the momentum and position operators of the resonator with effective mass $m_{eff}$ and frequency $\Omega_{m}$. And the cavity mode with the frequency of $\omega_{c}$ is written as the annihilation (creation) operator $\hat{a}$($\hat{a}^{+})$.
The interaction Hamiltonian $\hat{H}_{int}$ is expressed as $\hat{H}_{int} =\hbar G \hat{x}\hat{a}^{+}\hat{a}$ which describes the interaction between the optical field and the mechanical field. $G=\omega_{c}/L$ denotes the coupling strength with $L$ being the cavity length.
$\hat{H}_{dirve}$ represents the Hamiltonian of the input field which consisted of the pump and probe terms as
\begin{align}
\begin{split}
& \hat{H}_{dirve} =i \hbar \sqrt{\kappa_{ex}}(\varepsilon_{1}\hat{a}^{+}e^{-i\omega_{1} t}-\varepsilon_{1}\hat{a}e^{i\omega_{1} t})\\
& \quad \quad +i \hbar \sqrt{\kappa_{ex}}(\varepsilon_{p}\hat{a}^{+}e^{-i\omega_{p} t}-\varepsilon_{p}\hat{a}e^{i\omega_{p} t}).
\end{split}
\end{align}
Here $\kappa_{ex}$ represents the rate of the external loss, $\omega_{1}$($\omega_{p}$) denotes the frequency of the pump(probe) light with amplitude $\varepsilon_{1}$ ($\varepsilon_{p}$), and here we have $\varepsilon_{1}=\sqrt{P_{1}/\hbar\omega_{1}}$ and $\varepsilon_{p}=\sqrt{P_{p}/\hbar\omega_{p}}$. $P_{l}$ and $P_{p}$ are the power of the pump and probe field, respectively.

In the frame rotating at the pump field frequency $\omega_{1}$, the Heisenberg-Langevin equations of the system could be written as
\begin{align}
& \frac{d\hat{a}}{dt}=(i\Delta-\frac{\kappa}{2}-i G \hat{x})\hat{a}+\sqrt{\kappa_{ex}}(\varepsilon_{1}+\varepsilon_{p}e^{-i\Omega t})+\sqrt{\kappa_{0}}\hat{s}_{vac},\\
& \frac{d^{2}\hat{x}}{dt^{2}}+\Gamma_{m}\frac{d\hat{x}}{dt}+\Omega_{m}^{2}=-\frac{\hbar G}{m_{eff}}\hat{a}^{+}\hat{a}+\hat{F}_{th},
\end{align}
where $\Delta=\omega_{1}-\omega_{c}$ and $\Omega=\omega_{p}-\omega_{1}$. Here we introduced the decay rate of the cavity field ($\kappa$) and mechanical oscillator ($\Gamma_{m}$) classically. The quantum noise of the mechanical oscillator and cavity are denoted as $\hat{F}_{th}$ and $\hat{s}_{vac}$, with $\langle \hat{s}_{vac}(t), \hat{s}_{vac}^{+}(t') \rangle=\delta(t-t')$, $\langle \hat{s}_{vac}(t) \rangle=0$, $\langle\hat{F}_{th}(t) \hat{F}_{th}^{+}(t') \rangle=\Gamma_{m}$, and $\langle \hat{F}_{th}(t) \rangle=0$. In this work, the mean field approximation works, viz., $a(t)\equiv\langle\hat{a}(t)\rangle$, $a^{*}(t)\equiv\langle\hat{a}^{+}(t)\rangle$ and $x(t)\equiv\langle\hat{x}(t)\rangle$. Based on this approximation, the mean value Heisenberg-Langevin equations can be written in the $c$-number form as
\begin{align}
& \frac{da}{dt}=(i\Delta-\frac{\kappa}{2}-i G x)a+\sqrt{\kappa_{ex}}(\varepsilon_{1}+\varepsilon_{p}e^{-i\Omega t})  \label{cnumber1};\\
& \frac{d^{2}x}{dt^{2}}+\Gamma_{m}\frac{dx}{dt}+\Omega_{m}^{2}=-\frac{\hbar G}{m_{eff}}a^{*}a  \label{cnumber2}.
\end{align}

\begin{figure}[htbp]
\centering
\includegraphics[width=0.5\textwidth]{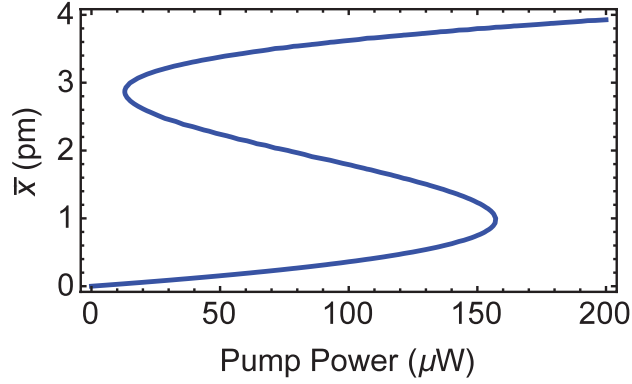}
\caption{(color online) Calculation results of the solutions of Eq.(\ref{steady}). Here we shows the affect of pump power $P_{1}$. We use $m_{eff}=2.0 pg$, $G=-485 GHz/nm$, $\Gamma_{m}/2\pi=35.0 kHz$,$\kappa/2\pi=50.0 MHz$, $\kappa_{ex}/2\pi=25.0 MHz$ and $\Omega_{m}=1.4 GHz$.}
\label{double}
\end{figure}

For the case that the probe field is much weaker than the pump field, the perturbation method is employed to address the issue of the above equations. The expression of the cavity field and the mechanical displacement under both the pump field and the probe field can be written as $a=\overline{a}+\delta a$ and $x=\overline{x}+\delta x$, and the steady state solution of the above equations could be solved as,
\begin{equation} \label{steady}
 \overline{a}=\frac{\sqrt{\kappa_{ex}}a_{in}}{-i\overline{\Delta}+\kappa/2},
 \quad \overline{x}=\frac{\hbar G |\overline{a}|^{2}}{m_{eff}\Omega_{m}^{2}},
\end{equation}
where the $\overline{\Delta}=\omega_{1}-(\omega_{c}+G \overline{x})$ represents the effective detuning of cavity resonance. Consider the situation where the pump laser is tuned close to the lower sideband, i.e., $\overline{\Delta}\approx-\Omega_{m}$, and we choose the system parameters as [40], $m_{eff}=2.0 pg$, $G=-485 GHz/nm$, $\Gamma_{m}/2\pi=35.0 kHz$, $\kappa/2\pi=50.0 MHz$, $\kappa_{ex}/2\pi=25.0 MHz$ and $\Omega_{m}=1.4 GHz$, the displacement $\overline{x}$ exhibits the bistable properties along with the variances of the control field which is shown in Fig.\ref{double}. The pump wavelength is chosen to be $\lambda=530nm$, we find that when $ P_{1}<300\mu W$, the displacement of mechanical oscillation is $\overline{x}<5 pm$ and $G \overline{x}\ll \Omega_{m}$. When $P_{1}<10 \mu W$ or $P_{1}>150 \mu W$, only one solution of the equation exists. Here in this study, we choose the pump power as $P_{1}=7.3 \mu W$ in order to avoid the bistable effect.


Considering the perturbation part, we use the following ansatz as
\begin{align}
\delta \hat{a}(t)¡¡& =A^{-}e^{-i\Omega t}+A^{+}e^{+i\Omega t}, \label{ansatz1}\\
\delta \hat{a}^{*}(t) & =(A^{+})^{*}e^{-i\Omega t}+(A^{-})^{*}e^{+i\Omega t}, \label{ansatz2}  \\
\delta x(t) & =X e^{-i\Omega t}+ X^{*}e^{+i\Omega t}. \label{ansatz3}
\end{align}
The radiation-pressure force will oscillate at the beat frequency $\Omega=\omega_{p}-\omega_{l}$ with the simultaneous presence of
pump and probe fields. When the beat frequency($\Omega$) is closed to the frequency $\Omega_{m}$, the mechanical mode starts to oscillate coherently which induces the stokes ($A^{+}$) and anti-Stokes scattering of light from the strong intracavity field. By substituting Eqs.(9-11) into Eq.(\ref{cnumber1}) and Eq.(\ref{cnumber2}), and the probe frequency $\omega_{1}+\Omega$ term, we can get the equations of $A^{-}$ and $A^{+}$ as
\begin{align}
(-i(\overline{\Delta}+\Omega)+\kappa/2)A^{-} & = -iG\overline{a}X+\sqrt{\kappa_{ex}}s_{p}, \\
(i(\overline{\Delta}-\Omega)+\kappa/2)(A^{+})^{*} & =iG\overline{a}X, \\
m_{eff}(\Omega_{m}^{2}-\Omega^{2}-i\Gamma_{m}\Omega) & =-\hbar G\overline{a}(A^{-}+(A^{+})^{*}).
\end{align}
The solution of these equation could be expressed as
\begin{align}
(A^{+})^{*} & = -\frac{if(\Omega_{m})}{-i(\overline{\Delta}+\Omega)+\kappa/2+2\overline{\Delta}f(\Omega_{m})}\sqrt{\kappa_{ex}}s_{p}\quad ,\label{stokes}\\
A^{-} & =\frac{1+if(\Omega_{m})}{-i(\overline{\Delta}+\Omega)+\kappa/2+2\overline{\Delta}f(\Omega_{m})}\sqrt{\kappa_{ex}}s_{p} , \label{probe}
\end{align}
with
\begin{align}
f(\Omega_{m}) & =\hbar G^{2}\overline{a}^{2}\frac{\chi(\Omega_{m})}{i(\overline{\Delta}+\Omega)+\kappa/2} ,\\
\chi(\Omega_{m}) & =\frac{1}{m_{eff}}\frac{1}{\Omega_{m}^{2}-\Omega^{2}-i\Omega \Gamma_{m}} .
\end{align}
Exploiting the input-output relation of the cavity, the transmission spectrum of the Stokes($t^{+}$) field and OMIT spectrum ($t^{-}$) could be written as
\begin{eqnarray}
t^{+}&=& t_{2\omega_{1}-\omega_{p}}=\frac{-if(\Omega_{m})\kappa_{ex}}{-i(\overline{\Delta}+\Omega)+\kappa/2+2\overline{\Delta}f(\Omega_{m})}, \\\label{t+}
t^{-}&=&t_{\omega_{p}} =1-\frac{(1+if(\Omega_{m}))\kappa_{ex}}{-i(\overline{\Delta}+\Omega)+\kappa/2+2\overline{\Delta}f(\Omega_{m})}.
\label{t-}
\end{eqnarray}
For the OMIT signal using the homodyne detection, we can get [30]
\begin{equation}
t_{hom}\approx 1-t^{-}=\frac{[1+if(\Omega_{m})]\kappa_{ex}}{-i(\overline{\Delta}+\Omega)+\kappa/2+2\overline{\Delta}f(\Omega_{m})},
\end{equation}
here the intracavity probe power with frequency $2\omega_{l}-\omega_{p}$ could be detected from the homodyne signal.

\begin{figure}
\centering
\includegraphics[width=0.56\textwidth]{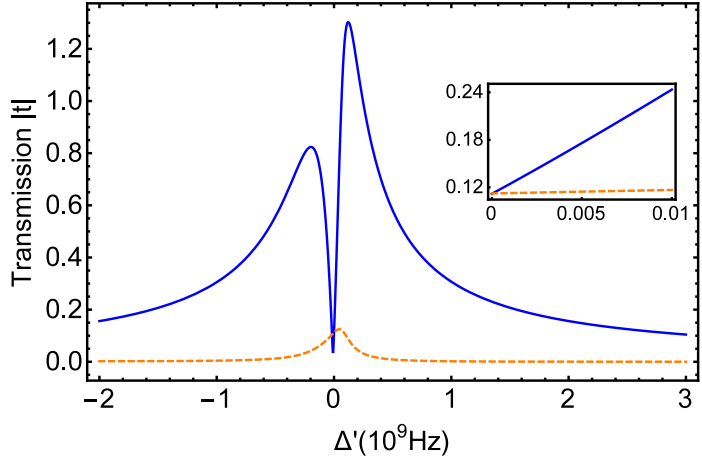}
\caption{The transmission the Stokes and homodyne signal with the parameters, $m_{eff}=2.0 pg$, $G=-485 GHz/nm$, $\Gamma_{m}/2\pi=35.0 kHz$, $\kappa/2\pi=100.0 MHz$, $\kappa_{ex}/2\pi=25.0 MHz$ and $\Omega_{m}=1.4 GHz$, the blue solid line shows the transmission of homodyne signal $t_{hom}$, and the orange dash line shows the transmission of the stokes field $t^{+}$  }
\label{StokesOMIT}
\end{figure}

In Fig.(\ref{StokesOMIT}), we plot the transmission spectrum of Stokes and  homodyne signal field with different $\Delta'$, where $\Delta'=\Omega-\Omega_{m}$. In the OMIT windows area, the Stokes filed cannot be neglected as shown in the subgraph of Fig.(\ref{StokesOMIT}). We find that the stokes is equal to the intensity of the homodyne signal. Then the homodyne signal intensity varies directly with the effective mass of the mechanical resonator (or strictly, decrease of the mechanical resonator's frequency). Then the change the effective mass of mechanical resonator could be mapped onto change of output signal intensity. We take the stokes signal as a reference in order to confront the decay in signal propagation. The Stokes signal and homodyne signal have a negligible frequency difference, simultaneously, they can create at almost the same time in the optical cavity. So they will undergo the same decay when the environment is linear and have frequency  irrelevant dissipation.

\section{Nanomechiancal mass sensing using OMIT}

 Here in this part, we give a detailed analysis of the mass sensing process based on the Stokes signals in nanomechanical system.

\subsection{nanomechiancal frequency shift and amplitude variation}\label{nanomechiancal}

In order to achieve mass sensing, we firstly consider the relationship between the output signals and the nanomechanical frequency shift. Here we suggest that the output Stokes and homodyne signals have a frequency irrelevant dissipation symbols by $\gamma(t)$, then the power of the detection can be written as
\begin{align}
t'_{hom}=\int t_{hom}e^{-i\gamma(t_{1})t_{1}}dt_{1}, \\
t'^{+}=\int t^{+}e^{-i\gamma(t_{2})t_{2}}dt_{2}.
\end{align}
Consider that the detuning between the Stokes signal and homodyne signal is twice of the mechanical resonator level, the dispersion could be ignored. Meanwhile, we can set $t_{1}=t_{2}$, for the wavefront of these two signals are always coincident.  The $t_{hom}$ and $t^{+}$ signal are time irrelevant and we can define the relative intensity($K_{st}$) as
\begin{equation}
\begin{split}
K_{st}&=\bigg|\frac{t'_{hom}}{t'^{+}}\bigg|=\bigg|\frac{\int t_{hom}e^{-i\gamma(t_{1})t_{1}}dt_{1}}{\int t^{+}e^{-i\gamma(t_{2})t_{2}}dt_{2}}\bigg| \\
&=\bigg|\frac{t_{hom}\int e^{-i\gamma(t)t}dt}{t^{+}\int e^{-i\gamma(t)t}dt}\bigg|=\bigg|\frac{t_{hom}}{t^{+}}\bigg|=\bigg|1+\frac{1}{if(\Omega_{m})}\bigg|.
\end{split}
\label{kst}
\end{equation}

\begin{figure}[htbp]
\centering
\includegraphics[width=0.6\textwidth]{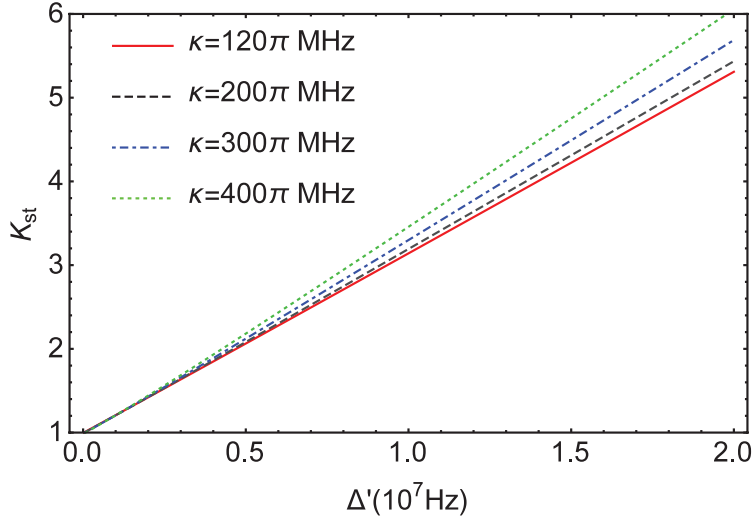}
\caption{(color online) The results of $K_{st}$ in Eq.(\ref{kst}). Here we shows the effect of pump power $P_{1}$, and we set the parameters as $m_{eff}=2.0 pg$, $G=-485 GHz/nm$, $\Gamma_{m}/2\pi=35.0 kHz$, $\kappa_{ex}/2\pi=50.0 MHz$ and $\Omega_{m}=1.4 GHz$.}
\label{reff}
\end{figure}

As shown in Eq.$(\ref{kst})$, by using the relative value of the homodyne and stokes signals, the mathematical form is also simplified by comparing with the form of homodyne signals and stokes signals. In Fig.$(\ref{reff})$, we plot the relative intensity $K_{st}$ with different frequency shift ($\delta\Omega_{m}=\Delta'$) under different decay rate ($\kappa$). Here we keep the system always under the critical coupling condition which means the coupling rate is related with the cavity decay rate, and we can conclude that there is a direct proportional relationship between the frequency shift($\delta\Omega_{m}$) and $K_{st}$. So the frequency can be easily read out from the variation of the relative intensity. According to the increment of the decay rate ($\kappa$), the relative intensity is more sensitive than the frequency shift.

\begin{figure}[htbp]
\centering
\includegraphics[width=0.6\textwidth]{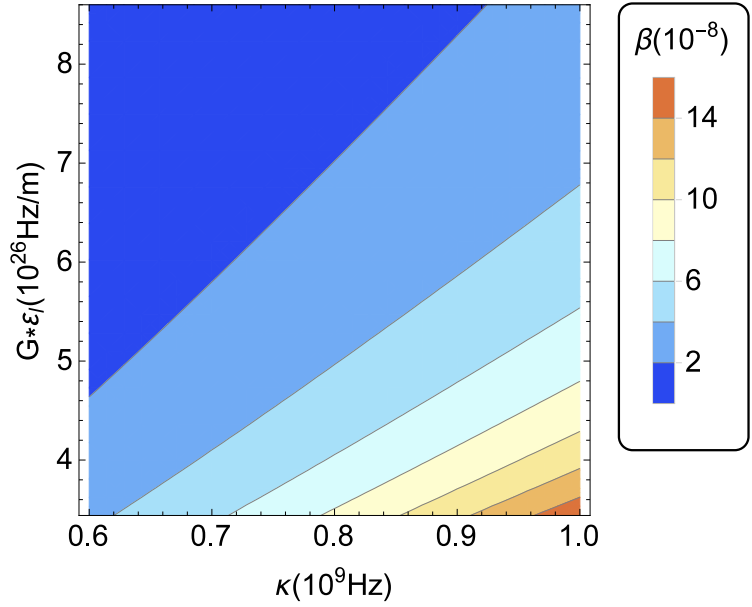}
\caption{(color online) The relative intensity of $\beta$ under different $\kappa$ and $G\epsilon_{l}$, with $m_{eff}=2.0 pg$, $\Gamma_{m}/2\pi=35.0 kHz$, $\kappa_{ex}/2\pi=50.0 MHz$, pump power $7\mu W$ with wavelength $\lambda=532nm$ and $\Omega_{m}=1.4 GHz$ .}
\label{massreff}
\end{figure}

\subsection{mess sensor under different parameters}\label{parameters}

As shown in Fig.$(\ref{StokesOMIT})$, the power of the Stokes signal is equal to the homodyne signal at $\Delta'=0$. Here we assume that the system is tuned to resonant with the frequency of the probe field. The frequency shift will happen due to the deposition of the nanoparticle on the mechanical system. Based on the relationship between the effective mass $m_{eff}$, spring constant $k$ and the fundamental resonance frequency [41] $\Omega_{m}=\sqrt{k/m_{eff}}$, the mass change($m_{d}$) of the mechanical resonator can be numerically solved as [42]
\begin{equation}
m_{d}=2\frac{m_{eff}}{\Omega_{m}}\delta\Omega_{m},
\label{sensing}
\end{equation}
And the relation between $m_{d}$ and $K_{st}$ could be described as
\begin{equation}
\begin{split}
K_{st}&=\frac{m_{eff}}{\hbar G^{2}\bar{a}^{2}}\bigg|[2\Omega_{m}+\frac{m_{d}\Omega_{m}}{2m_{eff}}-\frac{i\kappa}{2}]\\
&\times[2\Omega\frac{m_{d}\Omega_{m}}{2m_{eff}}-(\frac{m_{d}\Omega_{m}}{2m_{eff}})^{2}-i\Omega\Gamma_{m}]\bigg|.
\end{split}
\label{haha}
\end{equation}
Moreover, we define $\beta$as the slope of the $K_{st}$ value which can be express as
\begin{equation}
\beta=\frac{m\Omega^{2}(\gamma\kappa^{4}+12\gamma\kappa^{2}\Omega^{2}-4\kappa^{2}\Omega^{2}+16\Gamma\Omega^{2})}{G^{2}
\varepsilon_{l}^{2}\hbar\kappa_{ex}(\kappa^{2}+4\Omega^{2})\sqrt{\kappa^{2}+16\Omega^{2}}},
\label{daoshu}
\end{equation}
which reveals the relationship between the relative amplitude of the field and the mass change. Here we numerically simulated the value of $\beta$ under different $\kappa$ and $G\varepsilon_{l}$ in Fig.$(\ref{massreff})$. And what should be noticed is that, $\varepsilon_{l}$ determines the steady state of the system which is related with the pumping field. However, the destiny of the Stokes and  homodyne signals are mainly determined by the probe field. Here in Fig.$(\ref{massreff})$, we found that the slope of $\beta$ will be increased along with the increment of $\kappa$ or with the decrement of $G\varepsilon_{l}$.

\begin{figure}[htbp]
\centering
\includegraphics[width=0.6\textwidth]{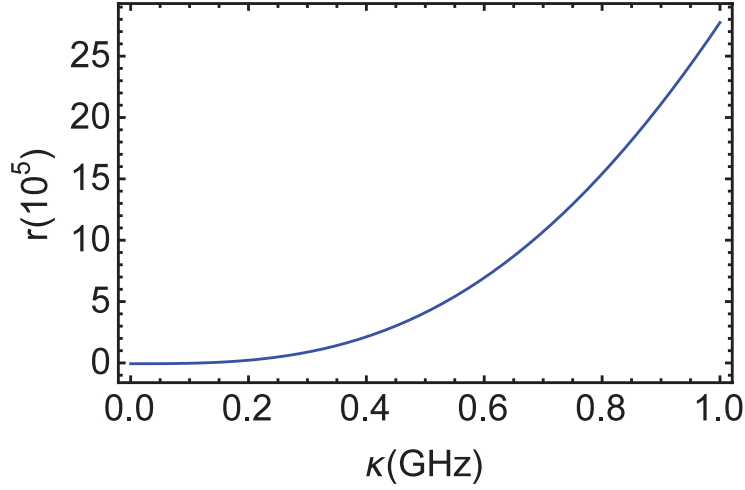}
\caption{(color online)$r$ under different $\kappa$ with $m_{eff}=2.0 pg$, $\Gamma_{m}/2\pi=35.0 kHz$, $\kappa_{ex}/2\pi=50.0 MHz$, pump power $7\mu W$ with wavelength $\lambda=532nm$ and $\Omega_{m}=1.4 GHz$ }
\label{G}
\end{figure}

\begin{figure}
\centering
\subfigure{
\begin{minipage}[b]{0.45\textwidth}
\includegraphics[width=\textwidth]{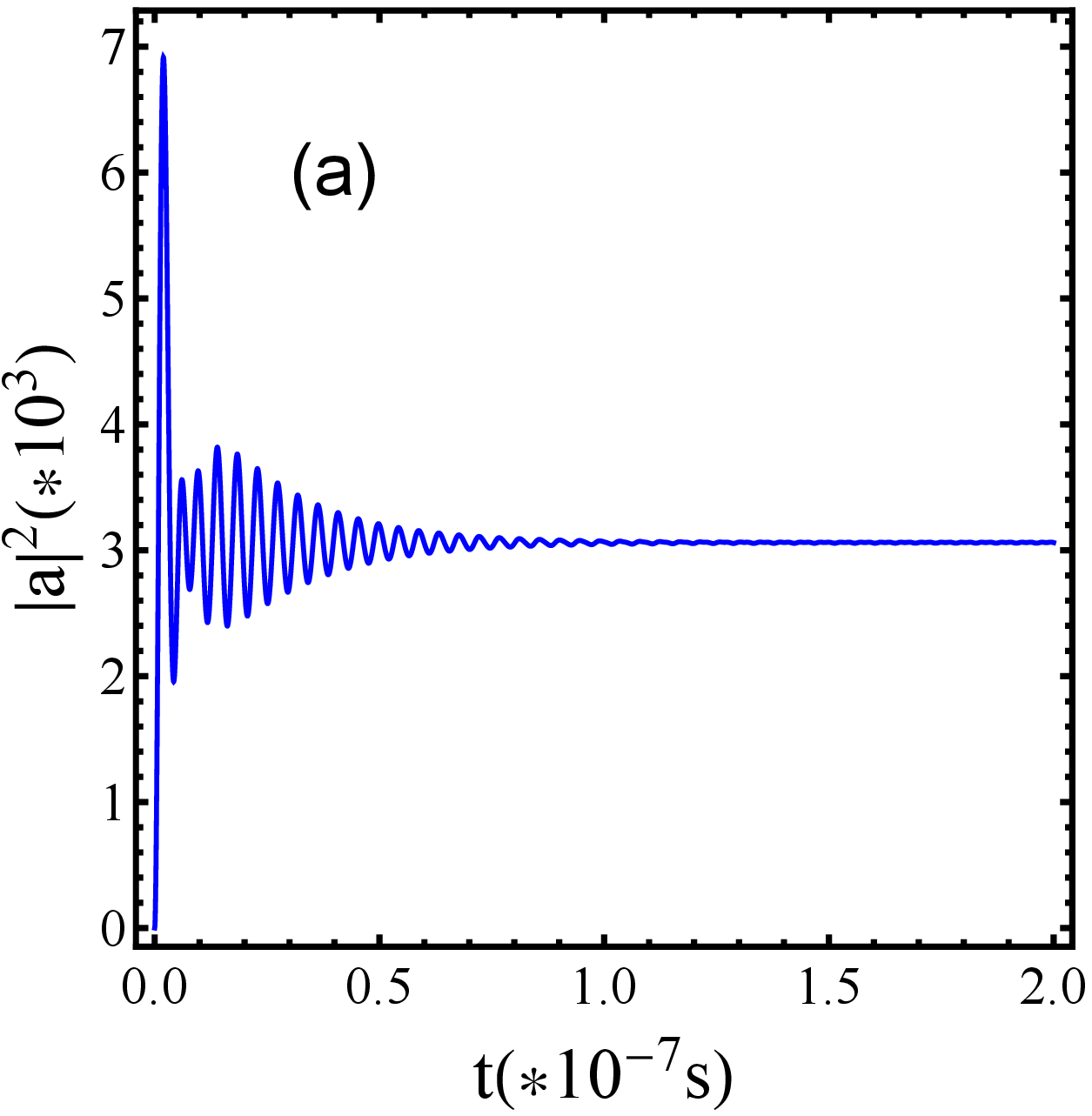}
\end{minipage}
}
\subfigure{
\begin{minipage}[b]{0.45\textwidth}
\includegraphics[width=\textwidth]{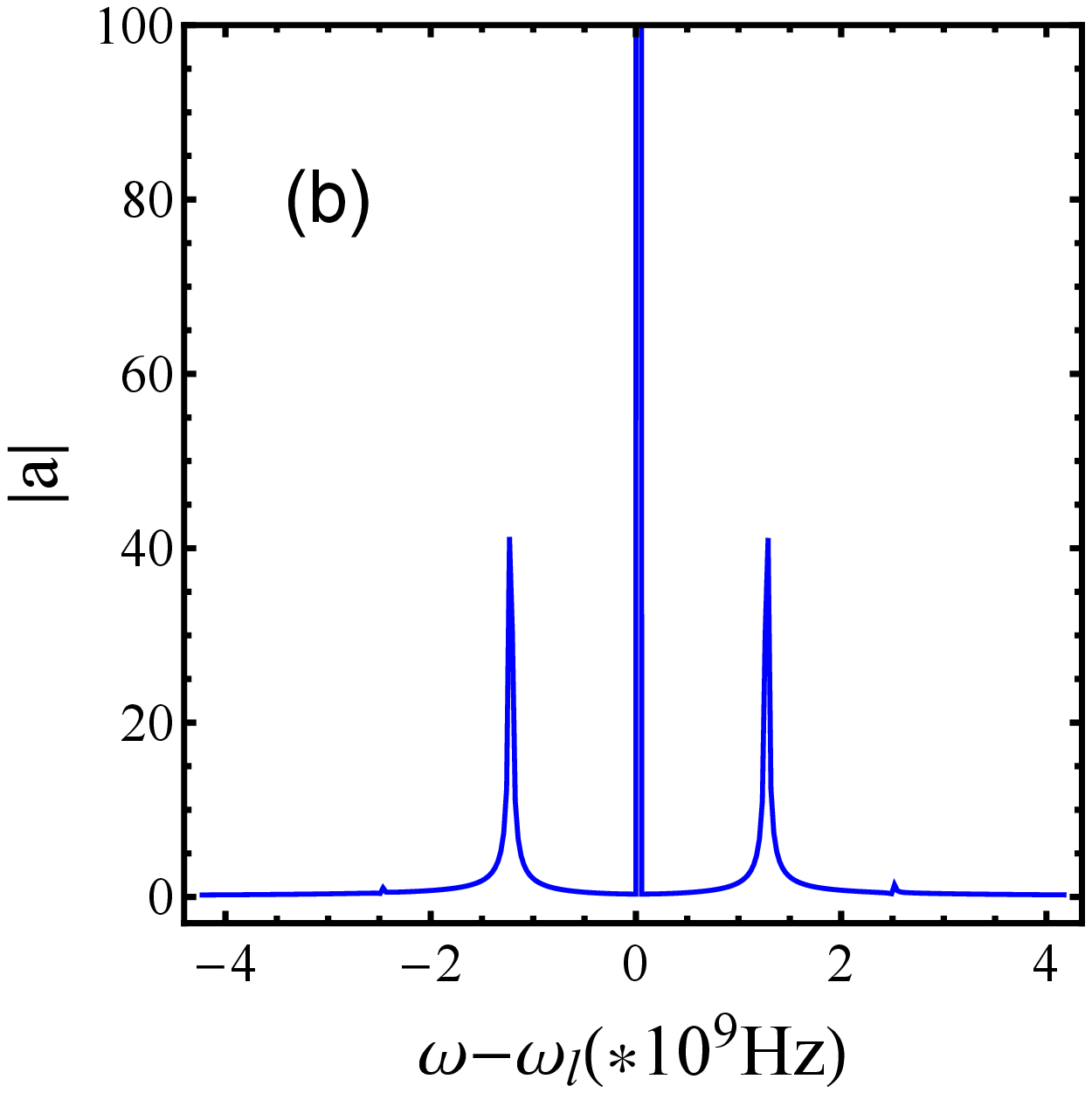}
\end{minipage}
}
\caption{The numerical simulation of the mass sensing spectrum on the frequency domain. Here we use the parameter as $m_{eff}=2.0 pg$, $G=-485 GHz/nm$, $\Gamma_{m}/2\pi=35.0 kHz$, $\kappa/2\pi=100.0 MHz$, $\kappa_{ex}/2\pi=25.0 MHz$ and $\Omega_{m}=1.4 GHz$. The pump power $P_{pu}=7.3\times10^{-6}W$, The probe power $P_{pr}=7.3\times10^{-9}W$.}
\label{simulation}
\end{figure}

In order to ensure the system works in the direct proportional area, the second order term should be much smaller than the first order term ($r=K'_{st}/K''_{st}\gg1$). We find that term $G\varepsilon_{l}$ will have no affect on this requirement, but $\kappa$ works. In Fig(\ref{G}), we plot the relationship between the first parameter of $r$ and $\kappa$.  This graph shows the direct proportional relationship is maintained when the decay rate($\kappa$) is greater than $0.1GHz$.

\subsection{The frequency spectrum of mass sensing}\label{simu}

Here in this part, we will illustrate the measurement process using this system. The detailed operations could be described as follow: (1) The frequency of the pump field is tuned to locate in the transparency window of OMIT. (2) When the probe field is sent into the system, after reaching the steady state, the field amplitude of the stokes and homodyne signals should be readout and the relative amplitude could be calculated. (3) Mass change of the nano-resonator induced by nanoparticle will have a slight change on the intensity, and the variance of the relative amplitude will be detected. (4) Read out the added mass according to the relationship between $m_{d}$ and $K_{st}$ as shown in Fig(\ref{massreff}) or Eq.(\ref{haha}).

As shown in Fig.(\ref{simulation}), we plot the time domain evolution and the steady state frequency spectrum under the condition of the mechanical-resonator. Here we can find the system reach its steady state in about $70ns$ from Fig.(\ref{simulation})(a), which means the mass sensing can be achieved in a quiet short time. Also we can get the reference of zero point. The Stokes and homodyne has almost the same amplitude in this point. Based on the zero point value, we can get the added mass on the nanomechancial  according to the relative amplitude change.

 \begin{figure}
\centering
\subfigure{
\begin{minipage}[b]{0.45\textwidth}
\includegraphics[width=\textwidth]{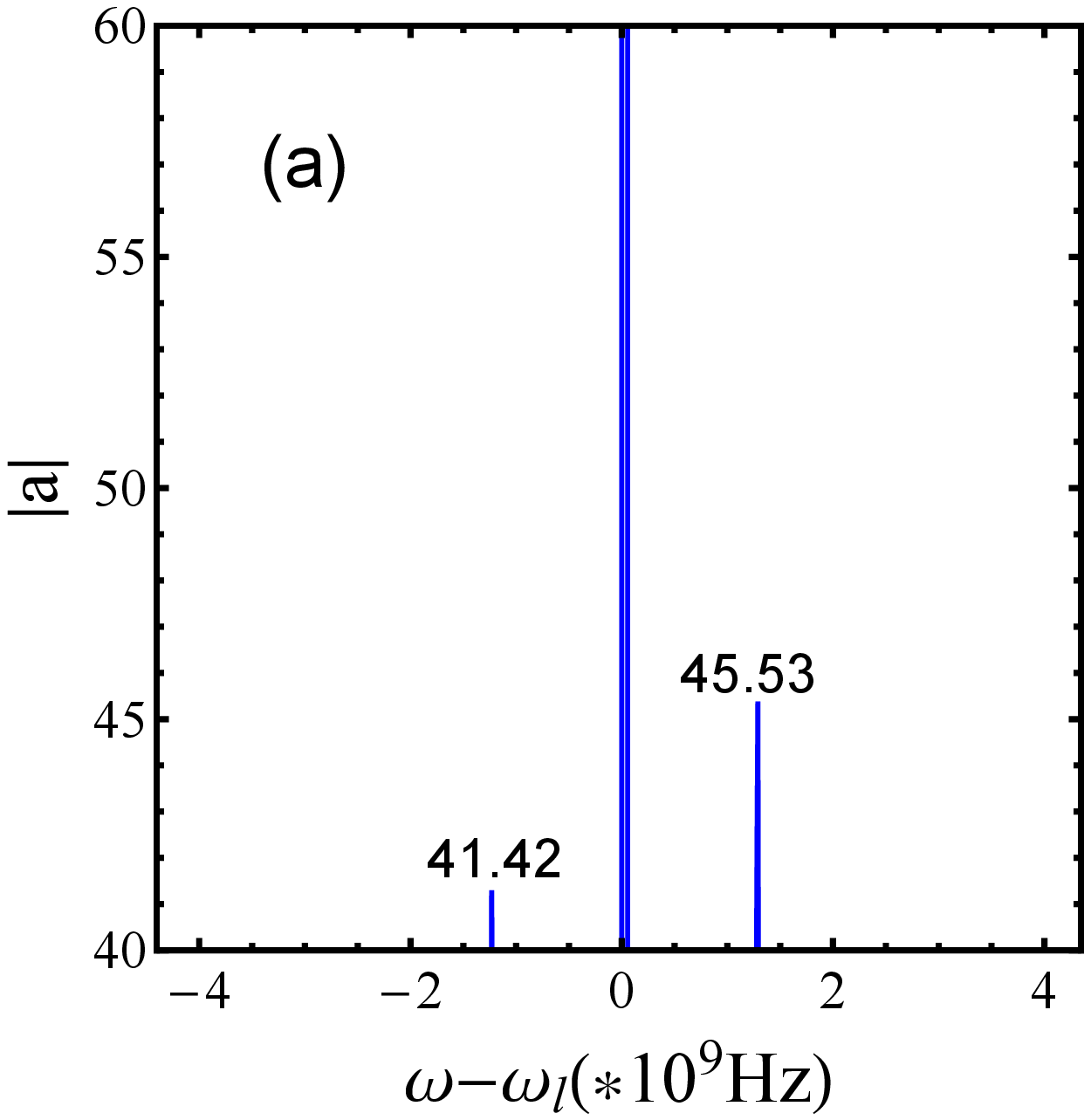}\\
\includegraphics[width=\textwidth]{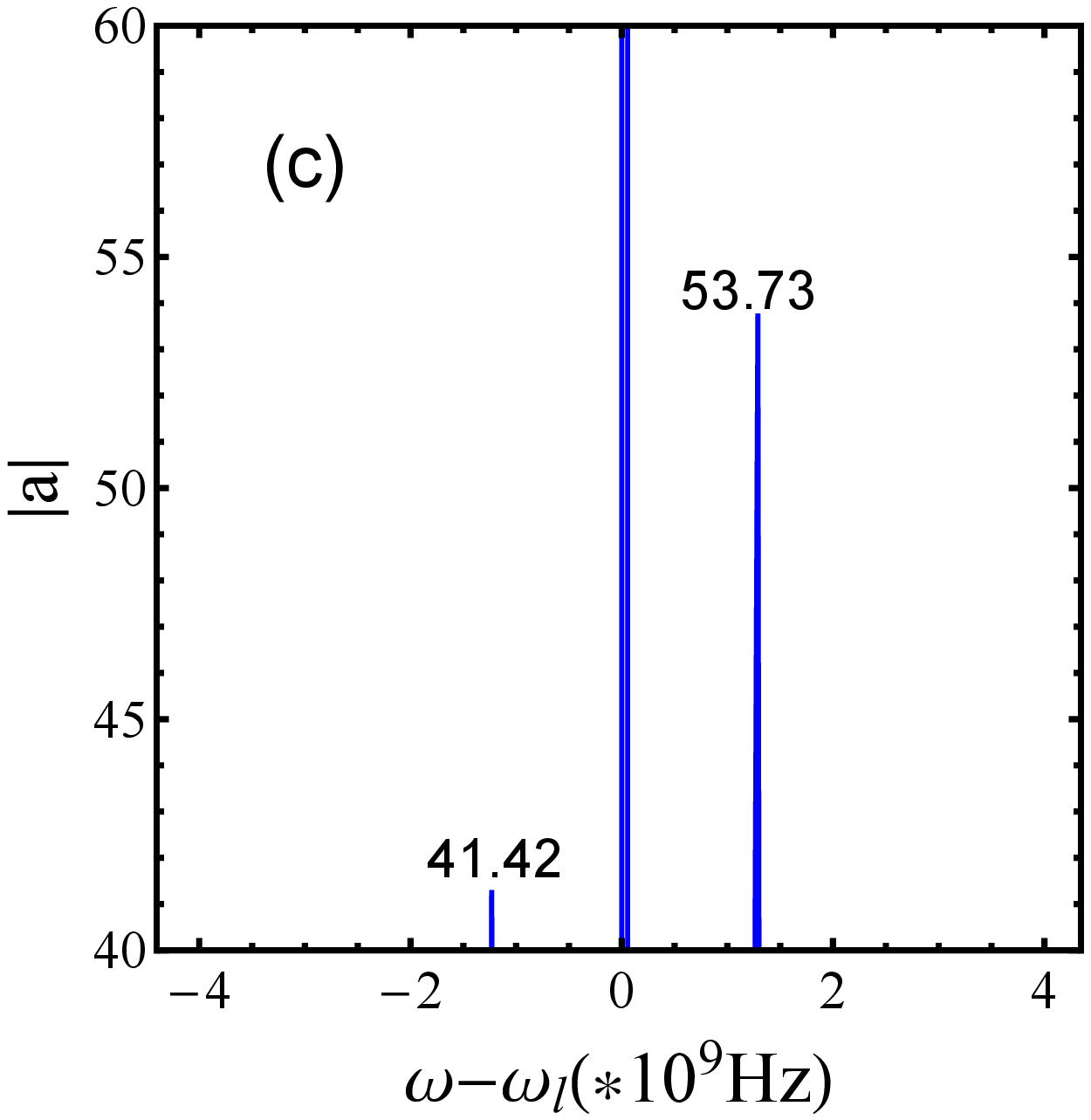}
\end{minipage}
}
\subfigure{
\begin{minipage}[b]{0.45\textwidth}
\includegraphics[width=\textwidth]{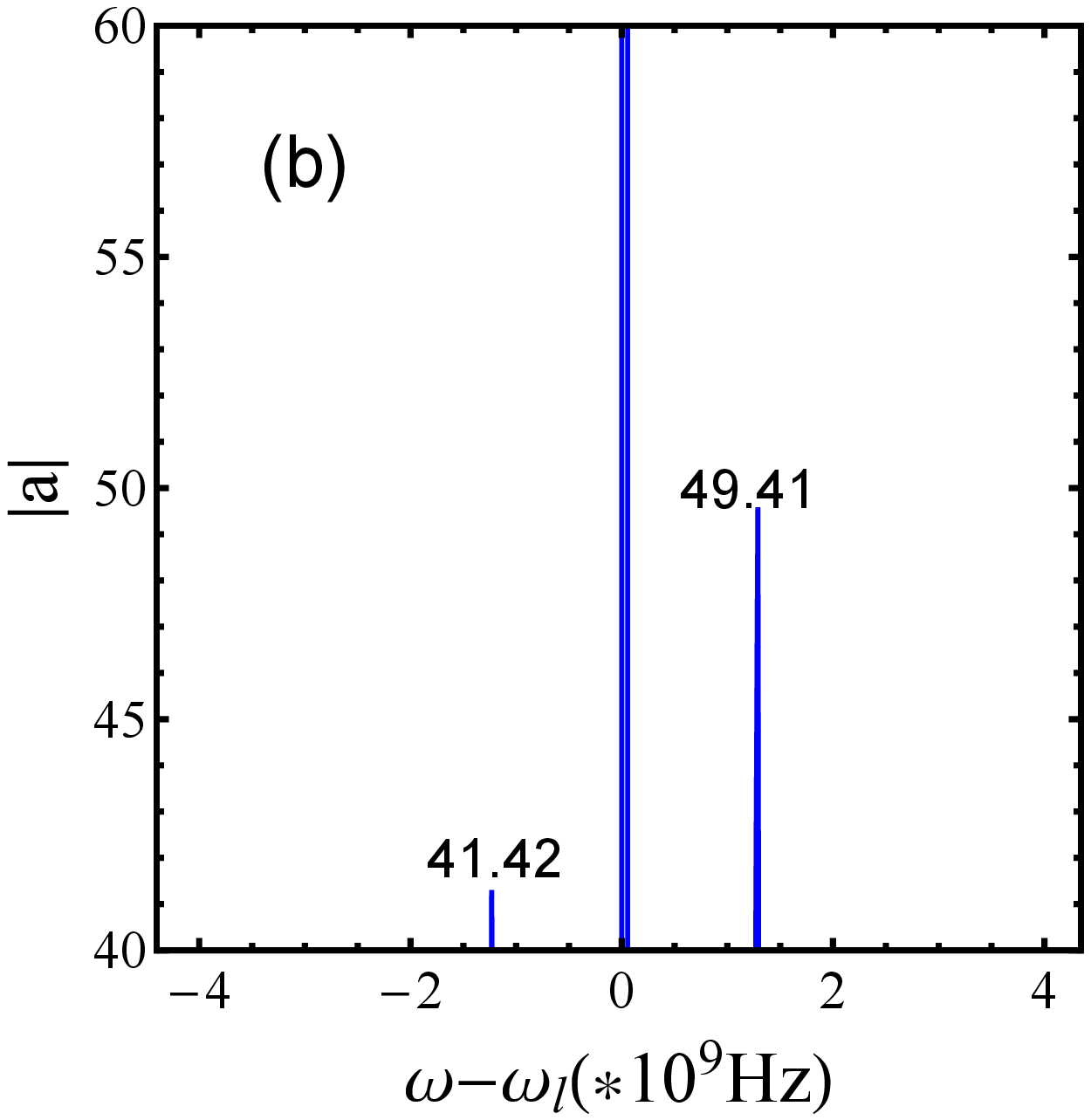}\\
\includegraphics[width=\textwidth]{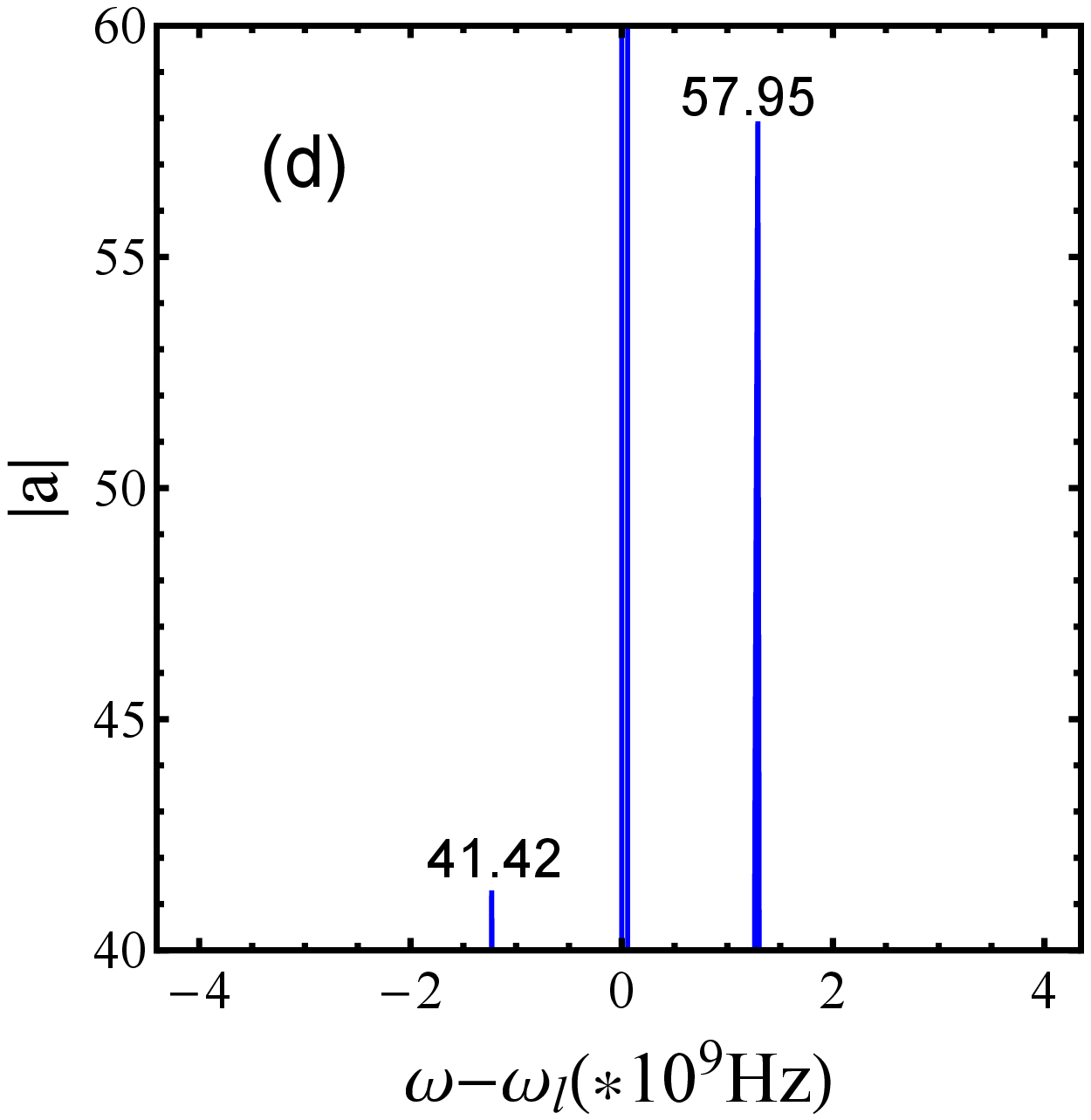}
\end{minipage}
}
\caption{The numerical simulation of the mass sensing spectrum on time domain and frequency domain. Here we use the parameter as $m_{eff}=2.0 pg$, $G=-485 GHz/nm$, $\Gamma_{m}/2\pi=35.0 kHz$, $\kappa/2\pi=100.0 MHz$, $\kappa_{ex}/2\pi=25.0 MHz$ and $\Omega_{m}=1.4 GHz$. The pump power $P_{pu}=7.3\times10^{-6}W$, The probe power $P_{pr}=7.3\times10^{-9}W$. (a) time evaluation of the cavity field of bare nano-resonator (b), frequency spectrum of the cavity field of bare nano-resonator.}
\label{sensing}
\end{figure}

As shown in Fig.(\ref{sensing}), we simulated the behavior of the output signal with the variation of the different nanoparticles with the mass of \{1.428, 2.857, 4.287, 5.671\}$fg$ on the nanomechanical resonators, and the output signal exhibits significant variations. As shown in Fig.(\ref{sensing}), we find that the homodyne signal shows different values of {45.53,49.41,53.75,57.95} and the stokes signal keeps unchanged with the value of 41.42. Based on the corresponding relationship between $K_{st}$ and $m_{d}$ shown in Fig.(\ref{massreff}) and Eq.(\ref{haha}). The sensing mass based on the simulation can be read as $\{1.323, 2.572,3.962,5.321\}fg$. And we may find there is a slight difference of 14\%. This error may be deduce to the nonlinear effect of the pump field increase the value of the homodyne and stokes field then the simulation value of the $K_{st}$ will be smaller. We can find the detection mass can be write as $\{1.410, 2.741, 4.223, 5.671 \}$. Then the maximal errors will reduce to 3\%. For the mended results, the error may comes from the discrete Fourier transform, the quasi-proportion relationship between $K_{st}$ and the added mass $m_{d}$, and the accuracy of the detection of the output frequency spectrum amplitude.

\section{Summary}

In summary, we have presented a stokes reference method to achieve mass sensing based on the intensity measurement of the output frequency spectrum. And we provide an expression to describe the added mass of the nano-resonator. We also discussed the performance of the mass sensing under different system parameters in the numerical simulation. And we find that at least the massing of femtogram sensing can be achieved even the cavity is Low-Q and the coupling is not strong enough.

%

\end{document}